\documentclass[a4paper,10pt,twoside]{cpc-hepnp}

\usepackage{multicol}
\usepackage{graphicx}
\usepackage{booktabs}
\usepackage{amssymb,bm,mathrsfs,bbm,amscd}
\usepackage[tbtags]{amsmath}
\usepackage{lastpage}

\usepackage{color}

\begin{document}

\fancyhead[co]{\footnotesize ZHANG Tong~ et al: FEL Polarization Control Studies on Dalian Coherent Light Source}


\title{FEL Polarization Control Studies on Dalian Coherent Light Source\thanks{Supported by Major State Basic Research Development Program of China (2011CB808300), and Natural Science Foundation of China (11175240 and 11205234)}}

\author{%
      ZHANG Tong$^{1,2}$
\quad DENG Hai-Xiao$^{1;1)}$\email{denghaixiao@sinap.ac.cn}%
\quad ZHANG Wei-Qing$^{3}$
\quad WU Guo-Rong$^{3}$\\
\quad DAI Dong-Xu$^{3}$
\quad WANG Dong$^{1}$
\quad YANG Xue-Ming$^{3}$
\quad ZHAO Zhen-Tang$^{1}$
}
\maketitle

\address{%
$^1$ Shanghai Institute of Applied Physics, Chinese Academy of Sciences, Shanghai 201800, China\\
$^2$ University of Chinese Academy of Sciences, Beijing, 100049, China\\
$^3$ State Key Laboratory of Molecular Reaction Dynamics,Dalian Institute of Chemical Physics, Chinese Academy of Sciences, Dalian, 116023, China 
}

\begin{abstract}
The polarization switch of a free-electron laser (FEL) is of great importance to the user scientific community. In this paper, we investigate the generation of controllable polarization FEL from two well-known approaches for Dalian coherent light source, i.e., crossed planar undulator and elliptical permanent undulator. In order to perform a fair comparative study, a one-dimensional time-dependent FEL code has been developed, in which the imperfection effects of an elliptical permanent undulator are taken into account. Comprehensive simulation results indicate that the residual beam energy chirp and the intrinsic FEL gain may contribute to the degradation of the polarization performance for the crossed planar undulator. And the elliptical permanent undulator is not very sensitive to the undulator errors and beam imperfections. Meanwhile, with proper configurations of the main planar undulators and additional elliptical permanent undulator section, circular polarized FEL with pulse energy exceeds 100 $\mu$J could be achieved at Dalian coherent light source.
\end{abstract}



\begin{multicols}{2}

\section{Introduction}

With the successful operation of the world's first two hard x-ray free electron laser (FEL) facilities, i.e., linear coherent light source~\cite{Emma_2010_LCLS_NP} and Spring-8 angstrom compact FEL~\cite{Ishikawa_2012_SACLA_NP}, high-gain single-pass FEL can work into hard x-ray regime, thus much smaller scientific field could be investigated for the first time~\cite{Barty_2012_LCLSexp_NP}. Currently, FEL community is on the stage to more sophisticated and determined schemes, e.g., in pursuit of compact design~\cite{Deng_2012_HarmonicXFELO_PRL,Huang_2012_TGU_PRL,Deng_2012_HarmonicXFEL_CPC}, fully coherence~\cite{Yu_2000_HGHG_Science,Jia_2008_EHGHG_APL,Stupakov_2009_EEHG_PRL,Xiang_2010_EEHG_PRL,Zhao_2012_SDUVEEHG_NP,Amann_2012_LCLSself-seeding_NP} and flexible polarization switch~\cite{Kim_2000_crossed,Geng_2010_crossed_NIMA,Geloni_2011_polar_DESYreport}.  Polarization is one of the fundamental optical properties of the light field. Circularly polarized lights with left or right spin could be used as the efficient probes for exciting chirality compounds. Taking the great advantages of FELs, powerful coherent radiation pulses with controllable polarization could be generated from the FEL facilities by various contraptions. Among which, crossed planar undulator (CPU) and elliptical permanent undulator (EPU) are the two most prevalent approaches.

CPU based approach was proposed by Kim~\cite{Kim_2000_crossed} and further extended~\cite{Geng_2010_crossed_NIMA,Ding_2008_crossed_PRSTAB,Li_2010_crossed_NIMA}  in the application of high-gain FEL, from which two identical planar undulators with orthogonal magnetic orientations are responsible for the two linear polarized FEL field along the crossed directions, i.e. $E_x$ and $E_y$, respectively. The phase between the two fields can be dynamically controlled by tuning a phase shifter between the two crossed undulators, thus the FEL with variable polarization states could be achieved. It is reported that CPU based approach has already experimentally realized on the storage ring~\cite{Litvinenko_2001_DukeOK_NIMA,Wu_2006_DukeOK_PRL}. Recently, CPU based polarization switch in seeded FEL and a proof-of-principle demonstration were proposed and under way at Shanghai Deep Ultra-violet Free-electron Laser SDUV-FEL facility~\cite{Zhao_2004_SDUV_NIMA,Zhang_2012_SDUVpolar_NIMA,Deng_2012_SDUVpolar_FEL2012}. Numerical investigation indicates that, highly circular polarized FEL light can be generated from a pair of CPU, in the presence of the beam energy chirp and the slippage effects. Moreover, by tuning the phase-shifter with a high repetition rate, fast helicity switching of FEL pulse could be obtained. EPU based approach could generate $E_x$ and $E_y$ simultaneously, and the phase difference between them could be adjusted with the subtle mechanical structure. Thus almost perfect polarized FEL could be produced from EPU, which have already been demonstrated on the Elettra storage ring~\cite{Spezzani_2011_TriesteringPolar_PRL} and FERMI FEL user facility~\cite{Allaria_2012_FermiFEL_NP}. However, EPU based approach is incapable of fast polarization switching due to the slow mechanical motion.

Dalian coherent light source (DCLS) is the first FEL user facility approved by government in China~\cite{Zhang_2012_DCLS_IPAC2012}, the baseline of which is high-gain harmonic generation  FEL~\cite{Yu_2000_HGHG_Science}. It is convinced that besides the capabilities of tuning wavelength from 50 to 150 nm, DCLS could also be an ideal facility to supply variable polarized FEL pulses by adding one undulator section downstream the main planar undulator line. In this paper, the two abovementioned approaches for polarized FEL generation, i.e., CPU and EPU were investigated thoroughly by utilizing the DCLS parameters, on the basis of the dedicatedly developed one-dimensional time-dependent FEL numerical code, so-called \textsc{Pelican}.

\section{Dalian coherent light source}

The construction of Dalian Coherent Light Source will begin early this year. It is designed to be working on the high-gain harmonic generation (HGHG) FEL principle. With the state-of-the-art optical parametric amplification (OPA) seed laser technique and variable gap undulator, DCLS could be lasing at extreme ultra-violet wavelength regime of 50 to 150 nm with the FEL pulse energy surpassing 100 $\mu$J. The machine parameters of DCLS can be found from Table~\ref{tab:DCLSparam}. 

\begin{center}
	\tabcaption{ \label{tab:DCLSparam} Designed parameters of DCLS}
	\footnotesize
	\begin{tabular*}{80mm}{l@{\extracolsep{\fill}}lll}
		\toprule
		Parameter & Symbol & Value & Unit \\
		\hline
		Beam energy & $E_b$ & $300$ & $\mathrm{MeV}$ \\
		Relative energy spread & $\eta$ &$1\times10^{-4}$ &  \\
		Normalized emittance & $\epsilon_n$ & $1-2$ & $\mathrm{mm} \cdot \mathrm{mrad}$ \\
		Peak current & $I_{pk}$ & $300$ & $\mathrm{A}$ \\
		Seed laser Wavelength & $\lambda_\mathrm{seed}$ & $240-360$ & $\mathrm{nm}$ \\
		Seed laser Width (FWHM) & $\tau_\mathrm{seed}$ & $1.0$ & $\mathrm{ps}$ \\
		Radiator period length  & $\lambda_r$ & $30$ & $\mathrm{mm}$ \\
		Radiator Parameter      & $a_r$ & $0.3-1.6$ & \\
		FEL wavelength & $\lambda_\mathrm{FEL}$ & $50-150$ & $\mathrm{nm}$ \\
		FEL Pulse Energy & $W_\mathrm{FEL}$ & $\geq 100$ & $\mu$J \\
		\bottomrule
	\end{tabular*}
\end{center}

For different radiation wavelengths~\cite{Liu_2013_Tunable_PRSTAB,Allaria_2012_FermiTunable_NJP}, one should tuning the wavelength of OPA and the gaps of the modulator and radiators so as to adapt the FEL resonant equation $\lambda_s = \lambda_u (1+a_u^2)/2\gamma^2$ with $\lambda_s$ the resonant wavelength, $\lambda_u$ the undulator period length and $a_u$ the normalized undulator parameter (see Fig.~\ref{fig:DCLSlayout}). Thus with the different frequency up-conversion harmonic number of HGHG (e.g. 2, 3, 4, 5), DCLS could emit FELs with continuously tunable frequency, which is truly useful to the specific FEL users.
\end{multicols}
\ruleup
\begin{center}
	\includegraphics[width=12cm]{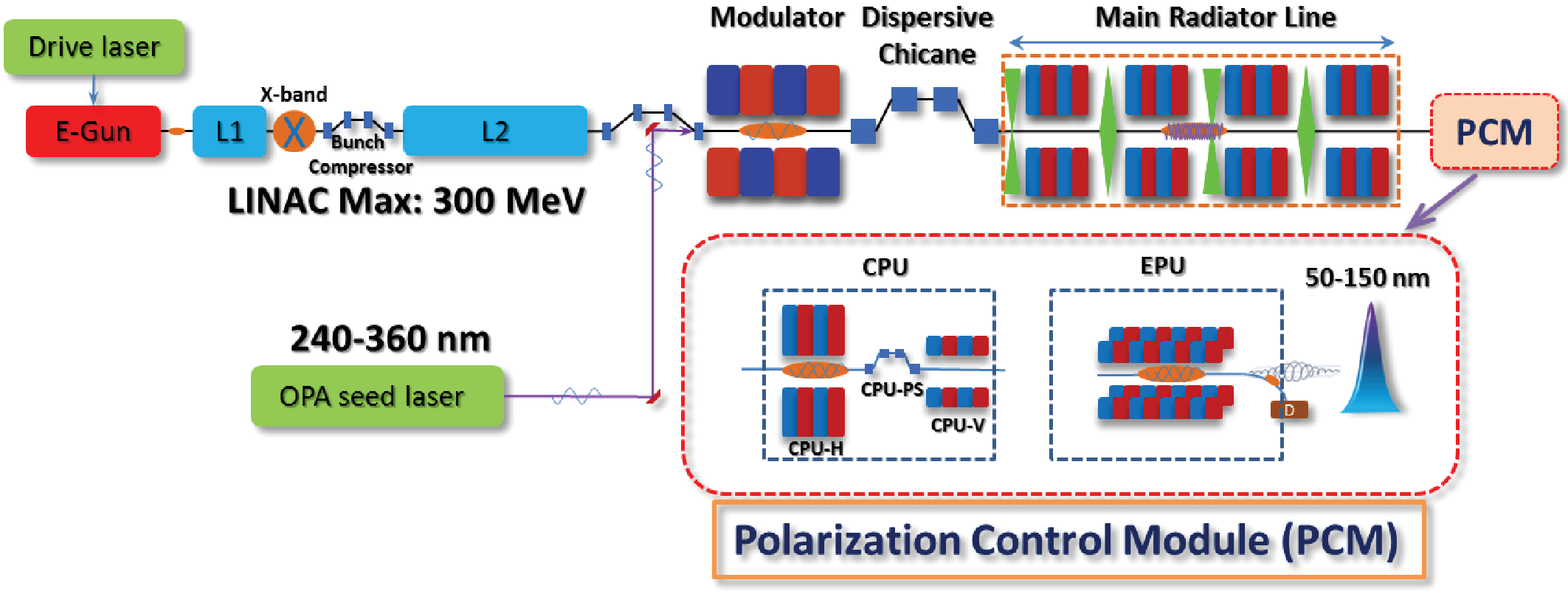}
	\figcaption{ \label{fig:DCLSlayout} Schematic layout of DCLS, the FEL polarization control module (PCM) has two options, CPU or EPU as shown in the figure.}
\end{center}
\ruledown \vspace{0.5cm}

\begin{multicols}{2}

\section{FEL code \textsc{Pelican}}

FEL polarization control at DCLS with CPU and EPU approaches has been carefully investigated with the one-dimensional time-dependent FEL code \textsc{Pelican}, with parallel computing enabled by default. \textsc{Pelican} is originally developed to figure out the EPU-related FEL polarization performance, which cannot be properly resolved in the publicly available FEL code, as far as our knowledge. The numerical model of \textsc{Pelican} is the coupled first-order FEL equation, written as~\cite{Saldin_2000_FELbook_Springer,Schmuser_2008_FELbook_Springer},
\end{multicols}
\ruleup	
\begin{eqnarray}
\label{eqn:eqn1}
	\frac{d\phi_\kappa}{dz} &=& 2 k_u \eta_\kappa, \kappa = 1,2,3 \cdots N \\
\label{eqn:eqn2}
	\frac{d\eta_\kappa}{dz} &=& -\frac{e}{m_e c^2 \gamma_r} \mathcal{R} \left\lbrace \left( \frac{\hat{K}\hat{E}\left(z,u_\kappa\right)}{2\gamma_r} - \frac{i \mu_0 c^2 \hat{j_1} \left(z,\xi_\kappa \right)}{\omega_s} \right) e^{i\phi_\kappa} \right\rbrace \\ 
\label{eqn:eqn3}	
	\frac{\partial \hat{E}\left(z,u\right)}{\partial z} &=& -\frac{\mu_0 c \hat{K}}{4\gamma_r}j_0\left(c_m\right) \frac{2}{N_m} \sum_{\kappa \in I_m} e^{-ik_s\xi_\kappa}, c_m = m \lambda_s
\end{eqnarray}
\ruledown \vspace{0.5cm}
\begin{multicols}{2}
With $\xi_\kappa = \left( \phi_\kappa+\pi/2\right)/k_s$, $k_s = 2\pi/\lambda_s$ and $u_\kappa = \left(1-\frac{1}{\overline{\beta}}\right)z+\frac{1}{\overline{\beta}}\xi_\kappa$,
where $\phi_\kappa$ is the so-called ponder-motive phase of the $\kappa-th$ electron particle, $k_u$ is the undulator wave number, $\eta_\kappa$ is the relative energy spread, $\gamma_r$ is the resonance beam energy, $\hat{E}$ is the FEL electric field, $\hat{j_1}$ is the first-order Fourier electron current density, $\xi_\kappa$ and $u_\kappa$ is the bunch coordinate in the electron frame and laboratory frame, respectively.

As the Eqs.~(\ref{eqn:eqn1})-(\ref{eqn:eqn3}) show,the universal longitudinal electron beam is treated as many slices separated by $m\lambda_s$, where m is a positive integer. The slippage effect between the light field and electron is also included. Besides, the FEL field $\hat{E}\left(z,u\right)$ could be substituted to be two orthogonal field vector $\hat{E}_x$ and $\hat{E}_y$ to resolve the helical undulator case, the undulator parameters $\hat{K}$ also should be corrected as $K_h = \hat{K}_p/JJ$ 
with $JJ = J_0\left(a_u^2/2\left(1+a_u^2\right)\right)-J_1\left(a_u^2/2\left(1+a_u^2\right)\right)$ 
the Bessel coupling factor of planar undulator.
	
\textsc{Pelican} can calculate polarization degrees of the FEL radiations for both CPU and EPU based principles, and can also simulate the undulator field error and phase error effect to the polarized FELs. Benchmarks have been carried out intensely between the widely-used three-dimensional FEL code \textsc{Genesis}~\cite{Reiche_1999_genesis_NIMA} and \textsc{Pelican}, the results shows good agreements. In the following sections, we use \textsc{Pelican} to study the polarized FELs generated from both CPU and EPU approaches.

\section{Crossed planar undulator case}

The crossed planar undulator is composed of three parts (see PCM in Fig.~\ref{fig:DCLSlayout}), two undulator modules with magnetic field in the vertical plane (e.g. $y-z$ plane, noted CPU-H) and orthogonal orientation ($x-z$ plane, noted CPU-V) and a small phase-shifter (CPU-PS). The phase-shifter is usually made of three electric dipoles, and stimulating the dipole with different current strengths will vary the path delay between the FEL electric field $E_x$ from CPU-H and $E_y$ from CPU-V, thus rotate the spin of circular polarization of the final combined FEL light downstream.
	
The CPU-H and CPU-V considered here is with the same total length of 1.5 m and period of 30 mm. The FEL polarization control is studied when the gaps of the 4 segments of main planar radiators of DCLS are opened up, thus the micro-bunched electron beam drift freely to the PCM and generate polarized FELs. Start-to-end (S2E) simulations have been comprehensively carried out to estimate the expected performances by the codes \textsc{Astra}~\cite{Flottmann_2008_astra}, \textsc{Elegant}~\cite{Borland_2000_elegant_LS287}, \textsc{Genesis} and \textsc{Pelican}.

It is reported in Ref.~\cite{Zhang_2012_SDUVpolar_NIMA} that the residual energy chirp of the electron beam could degrade the total polarization of the FEL radiations owing to the FEL slippage effect. At the FEL facilities like linear coherent light source, X-band structure is utilized to wipe off such residual energy chirp~\cite{Emma_2001_Xband_LCLS-TN}, thus two S2E cases i.e. with X-band on and off have been studies, respectively. The beam energy and current distributions along the bunch coordinate exit from the linear accelerator (LINAC) for each case are showed in Fig.~\ref{fig:s2ebeam}.
\begin{center}
	\includegraphics[width=8cm]{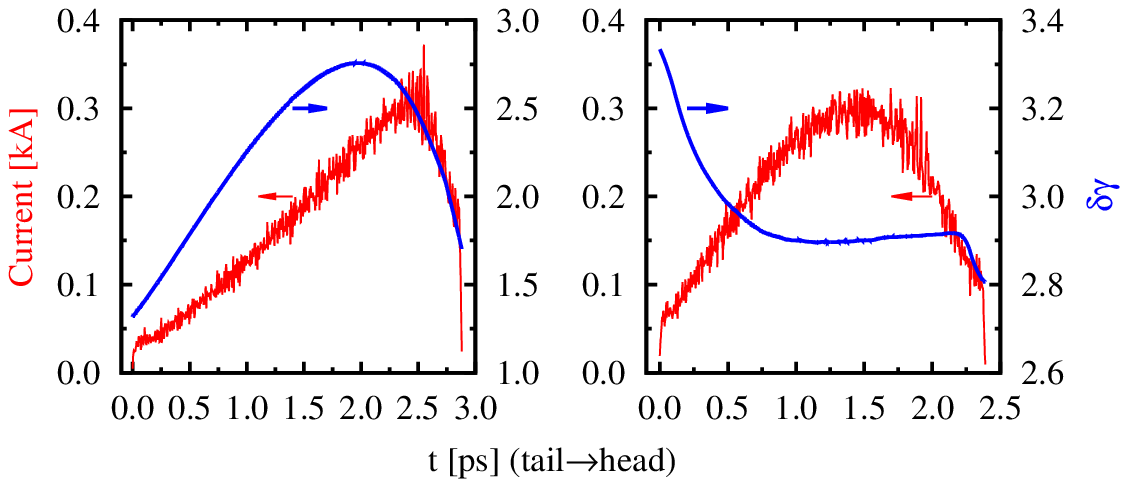}
	\figcaption{\label{fig:s2ebeam} Electron beam from the exit of LINAC, \emph{Left}:X-band off, \emph{Right}:X-band on.}
\end{center}

The S2E simulating process can be divided into three steps. Firstly, the particle description from \textsc{Elegant} is imported to \textsc{Genesis} for calculating the seed laser modulation on the electron bunch, usually with several thousands of slices since the entire bunch length is about 2-3 ps. Then \textsc{Pelican} takes care of the FEL process with $\tilde{E}_x$ and $\tilde{E}_y$ (the tilde sign '$\sim$' means complex quantity). Finally, by tuning the phase-shifter between CPU-H and CPU-V, the optimal circular polarized FELs is modeled. Fig.~\ref{fig:initbf} illustrates the initial bunching factors at the PCM entrance for different working wavelengths.
\begin{center}
	\includegraphics[width=7cm]{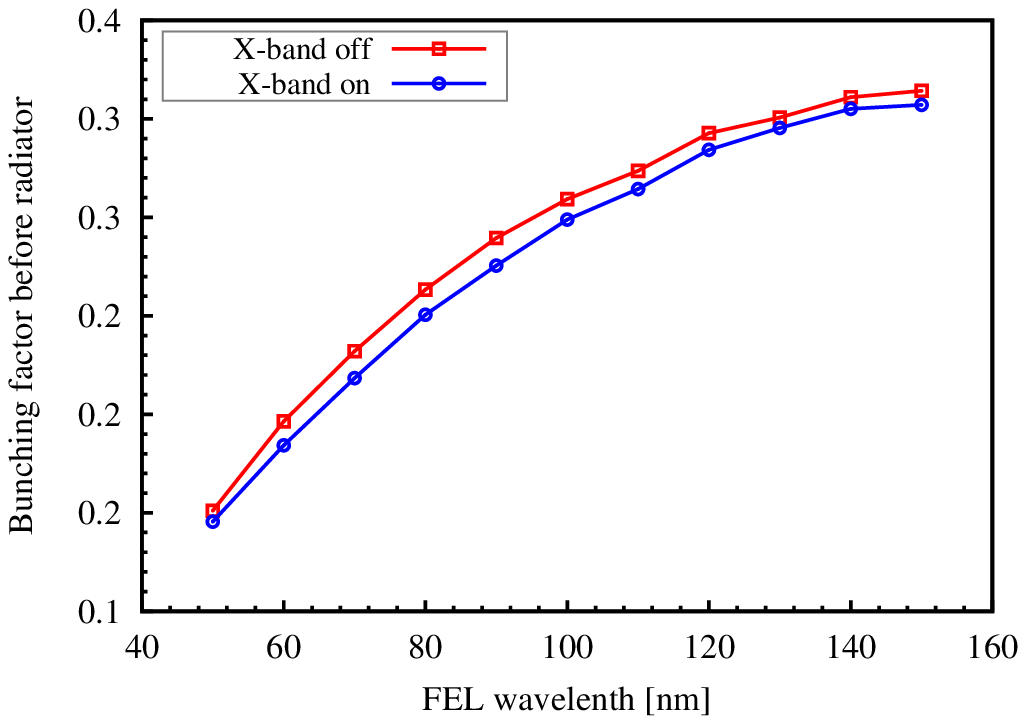}
	\figcaption{ \label{fig:initbf} Initial bunching factors at fundamental harmonic of the FEL radiation wavelengths in the spectral range of $50-150$ nm. \textcolor{red}{Red}   line: s2e case for X-band off,
	\textcolor{blue}{Blue} line: s2e case for X-band on.}
\end{center}

The horizontal linear polarized FEL, $E_x$ firstly produces when the micro-bunched electron beam traverse CPU-H with magnetic field in the $y$ direction. Simultaneously, longitudinal slippage between $E_x$ and electron beam happens, the consequence is that $E_x$ is naturally ahead $E_y$ by the total slippage length accomplished in CPU-H, thus the final received FEL radiations cannot be fully polarized. After passing through the phase-shifter, the electron beam continuously radiate in CPU-V where the vertical linear polarized FEL, $E_y$ is produced. According to the FEL physics, the former generated field $E_x$ is parallel with the undulator field of CPU-V, thus there is not any energy exchange between them. So the two orthogonal fields could be with almost the same field amplitude and constant phase difference, especially for the shorter wavelength regime in which the FEL gain length is much longer than the crossed undulator length. The characteristics of the CPU based polarized FELs have been simulated by \textsc{Pelican} for both beam case with X-band on and off, as shown in Fig.~\ref{fig:CPU_polar}.
\begin{center}
	\includegraphics[width=8cm]{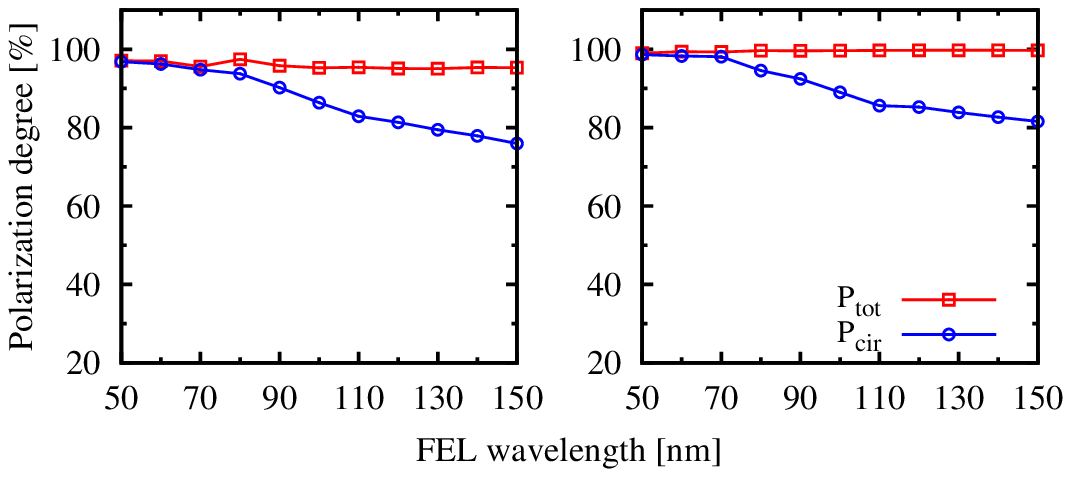}
	\figcaption{\label{fig:CPU_polar}
			 S2E simulated polarization degrees 
			 of the FEL radiations at the exit of CPU,
			 $P_{tot}$ and $P_{cir}$ is the total and circular 
		 	 polarization degree, respectively. 
		 	 \emph{Left}: X-band off, \emph{Right}: X-band on.}
\end{center}

Simulation results indicate that the total polarization always approach unity as the FEL wavelength increases, while the optimal circular polarization degree is dropping down. Accordingly, $P_{tot}$ is much more sensitive to the phase variation of the two linear polarized FEL pulses, while the field amplitude discrepancies greatly contribute to the $P_{cir}$ degradation. For the 1.5 m long crossed undulator here, the longer the radiation wavelength is, the more power FEL gains, which leads to a monotonously decrease tendency of $P_{cir}$ as the FEL wavelength increases. The helicity switch of the circular polarized FELs could be achieved by tuning the electric magnetic field of the small phase-shifter, as Fig.~\ref{fig:CPU_polarps_50_noX} shows.

\begin{center}
	\includegraphics[width=7cm]{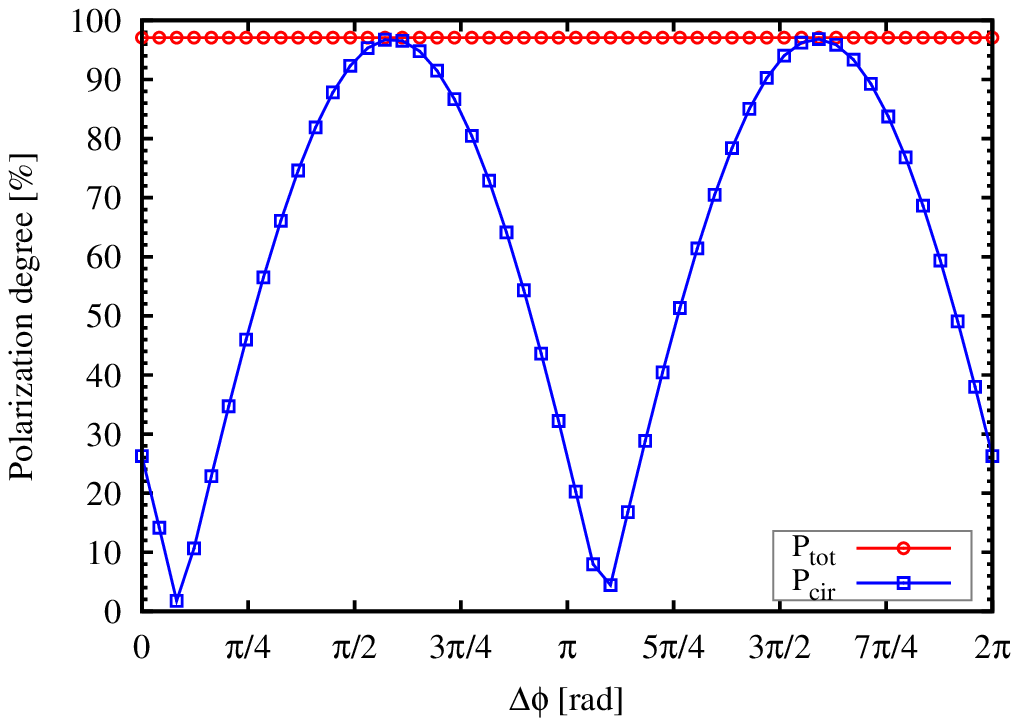}
	\figcaption{\label{fig:CPU_polarps_50_noX} 
			S2E simulation (X-band off) demonstration 
			of helicity switch of the 50 nm FEL radiations.}
\end{center}

\section{Elliptical permanent undulator case}
Elliptical permanent undulator is a conventional matured solution on the generation of circularly or elliptically polarized lights on the synchrotron radiation light source or FEL facility. Since the two crossed components of the electromagnetic fields are generated simultaneously in EPU, it is always nearly perfectly polarized. However the undulator errors of the EPU may break such perfection. Here we use \textsc{Pelican} to investigate the EPU-based FEL polarization manipulation cases for DCLS.

Under the same beam conditions as CPU mentioned, the micro-bunched electron beam just passes through one 3 m EPU section with the period length of 30 mm. It is truly that perfect circular polarized FELs are produced with an ideal EPU. Nevertheless the polarization degrees decrease as the phase error introduced. Fig.\ref{fig:EPU_phaseerr50_noX} shows that, with the RMS phase errors $\delta_{\phi_y}$ increases from 5 to 100 degree, the polarization perfection decreases as well, meanwhile the pulse energy of $E_y$ drops quickly. It is clearly that keeping the phase error within 20 degree in an EPU would achieve stable nearly fully circular polarized FEL radiation.
\begin{center}
	\includegraphics[width=8cm]{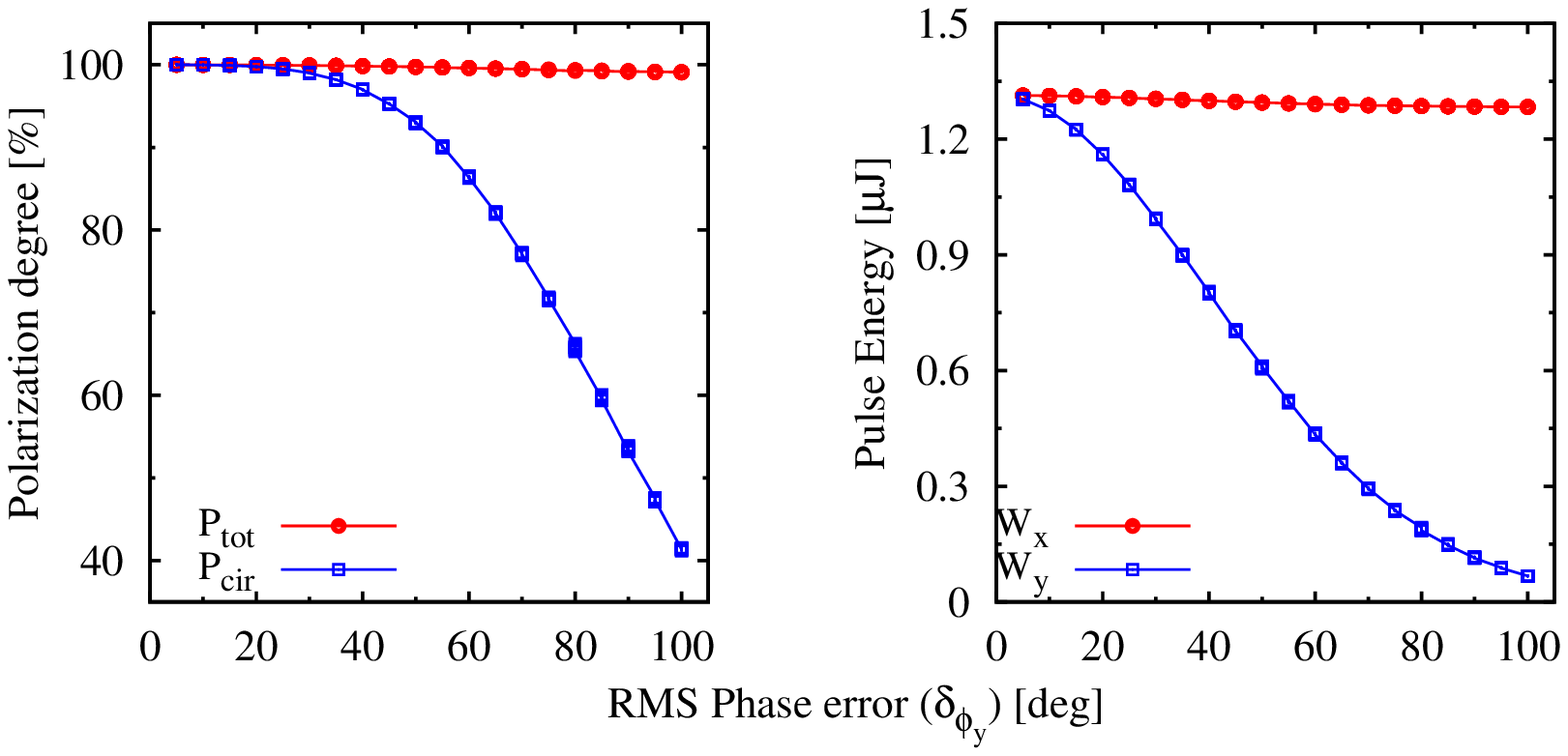}
	\figcaption{\label{fig:EPU_phaseerr50_noX} 
			 Relationship between polarization degrees 
		     and pulse energies with phase error at the wavelength
			 of 50 nm, here the X-band is off.}
\end{center}

One of the promising applications of EPU probably is the generation of powerful FELs with nearly perfect circular polarization. DCLS is designed to be producing linear polarized fully coherent FEL radiation with pulse energy surpassing 100 $\mu$J. While fully circular polarized powerful FELs also could be obtained by passing the partially radiated electron beam through the additional 3 m EPU module at the end of original main radiator line (as the Fig.~\ref{fig:DCLSlayout} shows).

After undergoing proper exponential gain in the main planar undulator, the generation of powerful highly circularly polarized FELs relies on the fact that the FEL power increases several times in the last EPU section. Here we demonstrate the case at the shortest wavelength of DCLS with $\lambda_s=50$ nm. The radiated electron bunch contributed linear polarized FEL pulse with the peak power of about 30 MW is sent into the last EPU to stimulate FEL gain at another orthogonal orientation to produce $E_y$ and simultaneously amplify $E_x$. Since the electron beam is already densely micro-bunched, within about 1-2 gain length, $E_x$ and $E_y$ grows significantly (see Fig.~\ref{fig:AEPU_pulse}). Fig.~\ref{fig:AEPU_polar} shows the polarization status evolution as the electron beam passes through the last EPU, with about 95\% circularly polarized FEL pulse is obtained finally; the pulse energy is nearly 130 $\mu$J while the output energy is only 2 $\mu$J in the case with only one EPU section is used.
\begin{center}
	\includegraphics[width=8cm]{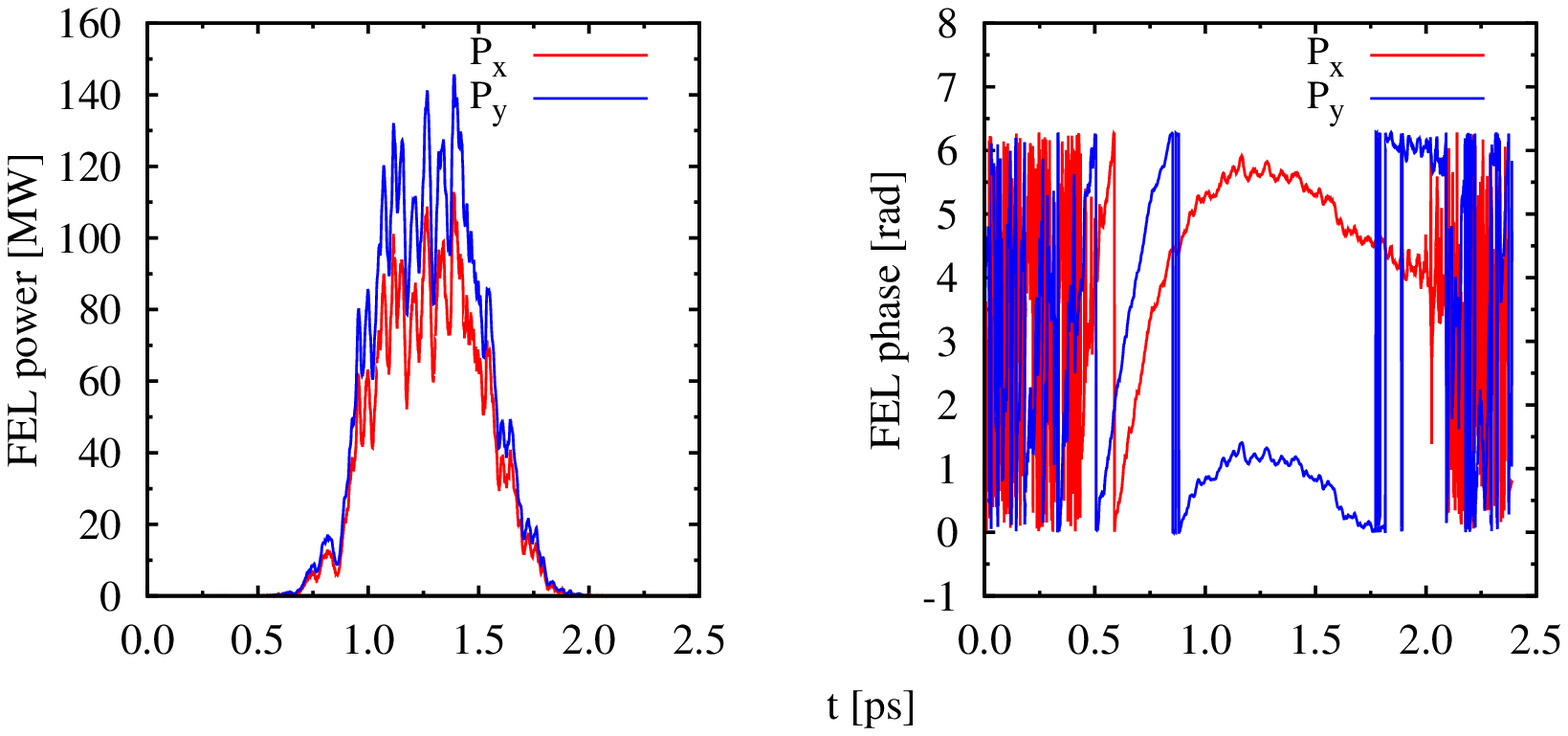}
	\figcaption{\label{fig:AEPU_pulse} FEL pulse from the last EPU}
\end{center}
\begin{center}
	\includegraphics[width=7cm]{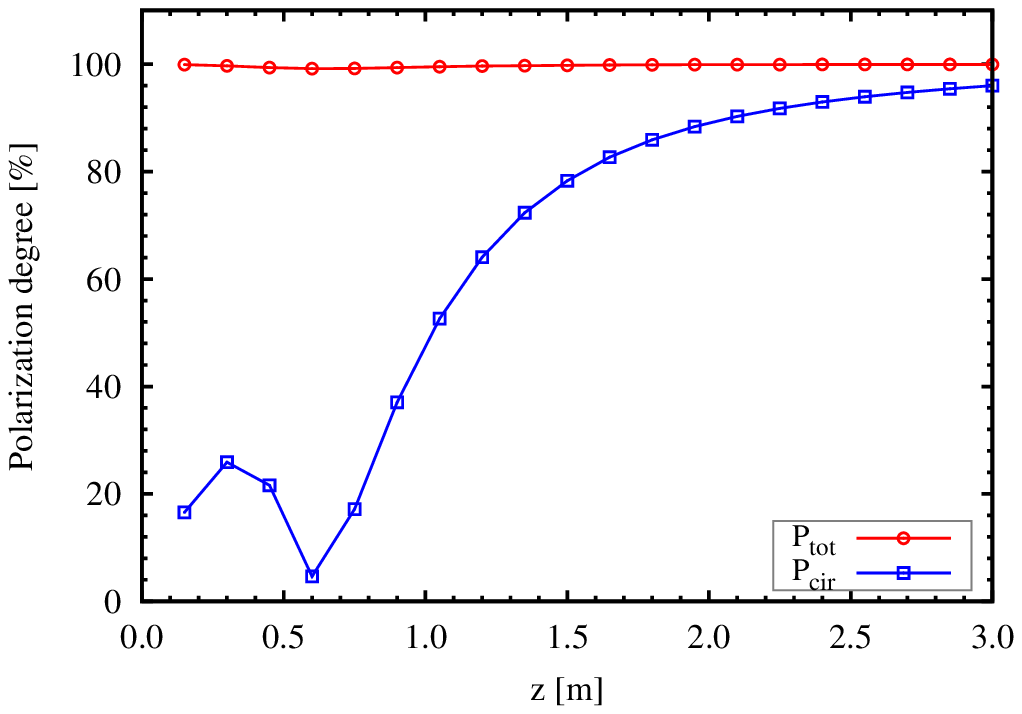}	
	\figcaption{\label{fig:AEPU_polar} 
			 Polarization status of the FEL pulse 
			 when the electron beam passes through the last EPU}
\end{center}

\section{Conclusions}
In this paper, the crossed planar undulator and elliptical permanent undulator based approaches of polarization controllable FEL pulse are comprehensively investigated for Dalian coherent light source. In order to fully understand and fairly compare the polarized FEL performance from CPU and EPU, a dedicated one-dimensional time-dependent FEL code has been developed. The start-to-end simulations present a bright future of the FEL polarization control at DCLS. Since the beam pulse repetition rate at DCLS is designed to be 50-100 Hz, the CPU approach may not be of quite benefit to the fast helicity switch, while it degrades the circular polarization performance at long wavelength due to the inequality of the field magnitude of two linearly polarized radiations, which is caused by the intrinsic FEL gain. Meanwhile, the EPU technique enlightens the potential of achieving powerful FEL pulses with nearly perfect circular polarization at DCLS from 50 to 150 nm, and it will definitely full-fledge the scientific research methodology. It is worth stressing that, for short-wavelength FEL machines such as NGLS~\cite{Corlett_2011_NGLS_PAC2011} where the beam repetition rate is of MHz order, CPU technique and the same idea, i.e., sandwiching two EPUs together with a phase-shifter still may be the ultimate novel approach.

\vspace{0.5cm}

\acknowledgments{The authors would like to thank Meng Zhang for providing tracked beam distribution and Chao Feng for helpful discussions.} 

\vspace{0.5cm}


\end{multicols}
\clearpage
\end{document}